\documentclass{aa}  
\usepackage[switch]{lineno} 

\usepackage{graphicx}
\usepackage{subcaption}
\usepackage{color}

\usepackage{txfonts}
\usepackage[colorlinks=true,linkcolor=blue,filecolor=blue,citecolor=blue,urlcolor=blue]{hyperref}
\usepackage{xspace}
\newcommand{\psr}{PSR\,J0742$-$2822\xspace}

\begin{document}

   \title{Glitch-induced pulse profile change of PSR J0742$-$2822 observed from the IAR}


   \author{E. Zubieta
          \inst{1,2}\fnmsep\thanks{E-mail: ezubieta@iar.unlp.edu.ar}
          \and
          F. Garc\'{i}a\inst{1,2}
          \and
          S. del Palacio\inst{1,3}
          \and
          C. M. Espinoza\inst{4,5}
          \and
          S. B. Araujo Furlan\inst{6,7}
          \and
          G. Gancio\inst{1}
          \and
          C. O. Lousto\inst{8,9}
          \and
          J. A. Combi\inst{1,2}
          \and
          E. G\"{u}gercino\u{g}lu\inst{10}
          }

   \institute{Instituto Argentino de Radioastronom\'ia (CCT La Plata, CONICET; CICPBA; UNLP),
              C.C.5, (1894) Villa Elisa, Buenos Aires, Argentina
         \and
             Facultad de Ciencias Astron\'omicas y Geof\'{\i}sicas, Universidad Nacional de La Plata, Paseo del Bosque, B1900FWA La Plata, Argentina
         \and
             Department of Space, Earth and Environment, Chalmers University of Technology, SE-412 96 Gothenburg, Sweden
         \and Departamento de F\'isica, Universidad de Santiago de Chile (USACH),  Av. V\'ictor Jara 3493, Estaci\'on Central, Chile.
         \and Center for Interdisciplinary Research in Astrophysics and Space Sciences (CIRAS), Universidad de Santiago de Chile.
         \and
             Instituto de Astronom\'\i{}a Teórica y Experimental, CONICET-UNC, Laprida 854, X5000BGR – Córdoba, Argentina
         \and
             Facultad de Matemática, Astronomía, Física y Computación, UNC. Av. Medina Allende s/n , Ciudad Universitaria, CP:X5000HUA - Córdoba, Argentina.
         \and
             School of Mathematical Sciences, Sciences Rochester Institute of Technology Rochester, NY 14623, USA
         \and
             Center for Computational Relativity and Gravitation, Rochester Institute of Technology, 85 Lomb Memorial Drive, Rochester, New York 14623, USA
        \and
            National Astronomical Observatories, Chinese Academy of Sciences, 20A Datun Road, Chaoyang District, Beijing 100101, China
            }
   \date{Received ...; accepted ...}

 
  \abstract
   {The radio pulsar PSR J0742$-$2822 is known to exhibit rapid changes between different pulse profile states that correlate with changes in its spin-down rate. However, the connection between these variations and the glitch activity of the pulsar remains unclear.}
   {We aim to study the evolution of the pulse profile and spin-down rate of PSR J0742$-$2822 in the period MJD 58810--60149 (November 2019 to July 2023), which includes the glitch on MJD 59839 (September 2022). In particular, we look for pulse profile or spin-down changes associated with the 2022 glitch.}
   {We observed PSR J0742$-$2822 with high cadence from the Argentine Institute of Radio astronomy (IAR) between November 2019 and July 2023. We used standard timing tools to characterize the times of arrival of the pulses and study the pulsar rotation, and particularly, the oscillations of $\dot \nu$. We also study the evolution of the pulse profile. For both of them, we compare their behavior before and after the 2022 glitch.} 
   {With respect to $\dot \nu$, we found oscillations diminished in amplitude after the glitch. We found four different components contributing to the pre-glitch $\dot \nu$ oscillations, and only one component after the glitch. About the emission, we found the pulse profile has two main peaks. We detected an increase in the $W_{50}$ of the total pulse profile of $\sim12\%$ after the glitch and we found the amplitude of the trailing peak increased with respect to the amplitude of the leading one after the glitch.}
   {We found significant changes in the pulse profile and the spin-down rate of PSR J0742$-$2822 after its 2022 glitch. These results suggest that there is a strong coupling between the internal superfluid of the neutron star and its magnetosphere, and that pulse profile changes may be led by this coupling instead of being led purely by magnetospheric effects.}


   \maketitle
%

\section{Introduction}
    Pulsars are neutron stars that present pulsed emission mainly in radio frequencies. The very high moment of inertia of pulsars provides them with a highly stable rotation \citep{2012MNRAS.427.2780H}. However, particularly in young pulsars, their stable rotation can be perturbed by timing noise and glitches. On the one hand, timing noise appears as a continuous and erratic pattern around the simple rotational evolution of the pulsar, and it is unclear whether it originates in the interior of the neutron star or due to alterations in the magnetosphere \citep{2009ApJ...700.1524M, 2010Sci...329..408L, 2017MNRAS.471.4827G, 2019MNRAS.489.3810P}. On the other hand, glitches are sudden changes in the rotation frequency of pulsars, generally accompanied by a change in the frequency derivative, which often returns to its pre-glitch value exponentially over different timescales, from one minute to several hundred days \citep{Zhou2022}. Glitches are considered to be provoked by the decoupling of the superfluid interior of the star from the solid crust \citep{1969Natur.224..673B, 2015IJMPD..2430008H, 2022MNRAS.511..425G}. Although this process is related to the dynamics inside the neutron star, glitches can also affect pulsar emission \citep{2013MNRAS.432.3080K,2022RPPh...85l6901A}.
    
    Investigating the impact of glitches in the spin evolution and the emission of the pulsar can reveal valuable information about the radiation processes occurring near the pulsar's surface and improve our understanding of its internal dynamics. Although nowadays at least 200 pulsars are known to present glitches\footnote{http://www.jb.man.ac.uk/pulsar/glitches.html} \citep{2011MNRAS.414.1679E, 2017A&A...608A.131F,2018IAUS..337..197M,2022MNRAS.510.4049B}, the relation between glitches, spin-down rate and emission changes in the radio band is a relatively understudied topic. To date, there are only a few reported emission changes clearly related to glitch events. 
   
   \cite{2010Sci...329..408L} showed that, for some pulsars, there was a strong correlation between spin-down rate and pulse-profile width, although no glitches were included in that study.  
   \cite{2011MNRAS.411.1917W} found extremely rare alterations in the pulse profile of PSR J1119$-$6227 preceded by glitches. \cite{2018MNRAS.478L..24K} showed that after a glitch in PSR B2035+36, the emission state transitioned to two different radiation modes, and there was a significant decrease in the pulse-profile width. \cite{2020Ap&SS.365...70S} showed that the separation between the peaks of the Crab pulsar's pulse profile displays an increasing trend with time. \cite{2020MNRAS.491.4634Y} reported that PSR J1048$-$5832 switches between a strong mode and a weak mode periodically. Another case is the one of PSR B1822$-$09, where two glitches were reported by \cite{2022ApJ...931..103L}, both associated with pulse profile changes. \cite{2023MNRAS.519...74Z} also reported a change in the emission profile of PSR J0738$-$4042 after a small glitch, which was also reported for PSR J1048$-$5832 by \cite{2024MNRAS.533.4274L}.

   In particular, \psr is known to switch between two magnetospheric states identified by different pulse profiles \citep{2010Sci...329..408L}, which correlates with changes in the spin-down rate, suggesting that timing noise can be driven by magnetospheric effects. However, \cite{2013MNRAS.432.3080K} found that the correlation between the spin-down rate and the pulse width increased strongly after a glitch, indicating that variations in the emission state might be linked to interactions between the neutron star interior and its magnetosphere. Finally, \cite{2021RAA....21...42D} and \cite{2022MNRAS.513.5861S} showed that there are different states of the pulsar characterised by the strength of the correlation between the pulse shape and spin-down rate and that not all the state changes in the spin-down rate are associated with glitches.

   So far nine glitches have been reported for \psr \citep{2022MNRAS.510.4049B}. The 2022 glitch (glitch \#9, \citep{2022ATel15622....1S,2022ATel15638....1Z,2024arXiv240514351G}) is an unique opportunity to study this pulsar, given that the relative frequency jump was three orders of magnitude higher than seven of the previous glitches (glitches \#1,\#2,\#3,\#4,\#5,\#6 and \#8) and 50 times greater than glitch \#7. In this work, we investigate the evolution of the pulse profile of \psr and spin-down rate along our data span between 2019 and 2023 ---including the 2022 glitch--- obtained from the Argentine Institute of Radioastronomy (IAR). 

   This paper is organized as following. In Sect.~\ref{sec:obs} we describe the pulsar monitoring campaign at IAR and the timing observations used for this work. In Sect.~\ref{sec: methods} we explain the methods used to track the pulsar rotation, characterize the spin-down rate behavior and study the pulse profile evolution. In Sect.~\ref{sec: results} we present the changes in the spin-down rate and the radio emission of the pulsar due to the glitch event, and finally in Sect.~\ref{sec:discussion} we discuss the implications of our results, with some concluding remarks in Sect.~\ref{sec:conclusions}.

\section{Observations}\label{sec:obs}

The IAR observatory is located near the city of La Plata, in Argentina. It has two 30-m single-dish antennas 120~m apart. The antennas "Carlos M. Varsavsky" and "Esteban Bajaja" are referred to as A1 and A2, respectively. Observations with A1 have a bandwidth of 112~MHz in one circular polarization at a central bandwidth of 1400~MHz, while observations with A2 have a bandwidth of 56~MHz in two circular polarizations at a central frequency of 1428~MHz \citep{2020A&A...633A..84G}. 
Since 2019, with the antennas of the IAR, our Pulsar Monitoring in Argentina\footnote{\url{https://puma.iar.unlp.edu.ar}} (PuMA) collaboration observes with high cadence a set of bright pulsars in the southern hemisphere \citep{2020A&A...633A..84G,2024RMxAC..56..134L}.

Among the monitored targets, there are a number of pulsars that have previously shown glitch activity, including \psr. Here, we analyze 408 observations of \psr obtained with A2 between MJD 58810 (November 23rd, 2019) and MJD 60149 (July 24th, 2023). The observations have a sample time of $146~\mu$s and a typical duration of 30~min, yielding a total of 204~h of observing time. We note that our observations are not flux-calibrated. This prevent us from studying the pulsar flux evolution.


\section{Methods}\label{sec: methods}

\subsection{Pulsar timing}

We processed all the observations with \textsc{PRESTO} \citep{2003ApJ...589..911R, 2011ascl.soft07017R}. First, we eliminated radio-frequency interferences (RFIs) with the task \textsc{rfifind} and folded the observations with the task \textsc{prepfold}. Then, we used the task \textsc{pat} in \textsc{PSRCHIVE} \citep{2004PASA...21..302H} to cross-correlate each profile with a template in order to calculate the Times of Arrival (TOAs). The template was obtained using the tool \textsc{psrsmooth} in \textsc{PSRCHIVE} on a pulse profile with a high signal-to-noise ratio (S/N).

We used TEMPO2 \citep{2006MNRAS.369..655H} to characterize the rotation of the pulsar by fitting the TOAs to a simple Taylor expansion:
\begin{equation}\label{eq:timing-model}
    \phi(t)=\phi+\nu(t-t_0)+\frac{1}{2}\dot{\nu}(t-t_0)^2+\frac{1}{6}\Ddot{\nu}(t-t_0)^3.
\end{equation}
where $t_0$ is the reference epoch, and $\nu$, $\dot\nu$ and $\ddot \nu$ are the rotation frequency and its first and second derivatives.

In the case of a glitch, the rotation rate suddenly increases. This change in the rotational phase provoked by the glitch can be modelled as \citep{2013MNRAS.429..688Y}:
\begin{multline}\label{eq:glitch-model}
    \phi_\mathrm{g}(t) = \Delta \phi + \Delta \nu_\mathrm{p} (t-t_\mathrm{g}) + \frac{1}{2} \Delta \dot{\nu}_\mathrm{p} (t-t_\mathrm{g})^2 + \\ 
    \frac{1}{6} \Delta \Ddot{\nu}(t-t_\mathrm{g})^3 + \sum_i \left[1-\exp{\left(-\frac{t-t_\mathrm{g}}{\tau^{i}_\mathrm{d}}\right)} \right]\Delta \nu^{i}_\mathrm{d} \, \tau_\mathrm{d}^{i}.
\end{multline}
Here, $\Delta \phi$ counteracts the uncertainty on the glitch epoch $t_\mathrm{g}$, and $\Delta \nu_\mathrm{p}$, $\Delta \dot{\nu}_\mathrm{p}$, and $\Delta \Ddot{\nu}$ are the step changes in $\nu$, $\dot\nu,$ and $\Ddot{\nu}$, at $t_\mathrm{g}$, respectively. Finally, $\Delta \nu_\mathrm{d}$ represents temporary increases in frequency that recovers in $\tau_\mathrm{d}$ days. From Eq.~\ref{eq:glitch-model}, we can define glitch sizes as

\begin{align}
    \frac{\Delta \nu_g}{\nu} &= \frac{\Delta \nu_\mathrm{p} + \sum_i \Delta\nu^{i}_d}{\nu} \\
    \frac{\Delta \dot\nu_g}{\dot\nu} &= \frac{\Delta \dot\nu_\mathrm{p} - \sum_i \Delta\nu^{i}_d/\tau^{i}_d}{\dot\nu},
\end{align}
We can also define the recovery factor for each recovery term as:

\begin{equation}
    Q_i=\frac{\Delta \nu^i_\mathrm{d}}{\Delta \nu_\mathrm{p}+ \sum_i \Delta \nu^i_\mathrm{d}},
\end{equation}

\noindent and the total recovery factor of the glitch:

\begin{equation}
    Q=\frac{\sum_i\Delta \nu^i_\mathrm{d}}{\Delta \nu_\mathrm{p}+ \sum_i \Delta \nu^i_\mathrm{d}}. 
\end{equation}

In addition, we characterized the oscillations of $\dot \nu$ with respect to the timing model. We used the Gaussian Process Regression (GPR) technique to fit the residuals with a square exponential covariance function \citep{2006gpml.book.....R,2022MNRAS.513.5861S} with the \textsc{gptools} library \citep{2015NucFu..55b3012C}. By analyzing the second derivative of this fitted function, we derived the evolution of $\dot \nu$ \citep{2013MNRAS.432.3080K}.
We then used the Lomb-Scargle spectral analysis tool \citep{1982ApJ...263..835S} to identify possible quasi-periodicities in the evolution of $\dot \nu$ and to evaluate the false alarm probability of the peaks using the \textsc{false\_alarm\_probability} method in the \textsc{astropy} library \citep{2013A&A...558A..33A}.

\subsection{Pulse profile evolution}\label{sec: gaussian}

We divided our data span into 26 intervals that exclude the glitch, which are at least 15 days long, and contain at least 15 TOAs. In each interval, we set $\ddot\nu=0$ and fit the values and uncertainties of $\nu$ and $\dot\nu$. We used these values to update the file header with the ephemeris of all pulse profiles within the region, to align their phases \citep{2024MNRAS.533.4274L}. Finally, we used the task \textsc{psradd} in \textsc{PSRCHIVE} to combine all profiles in the region, resulting in a single high S/N mean profile.

\begin{figure}[h]
    \includegraphics[width=\linewidth]{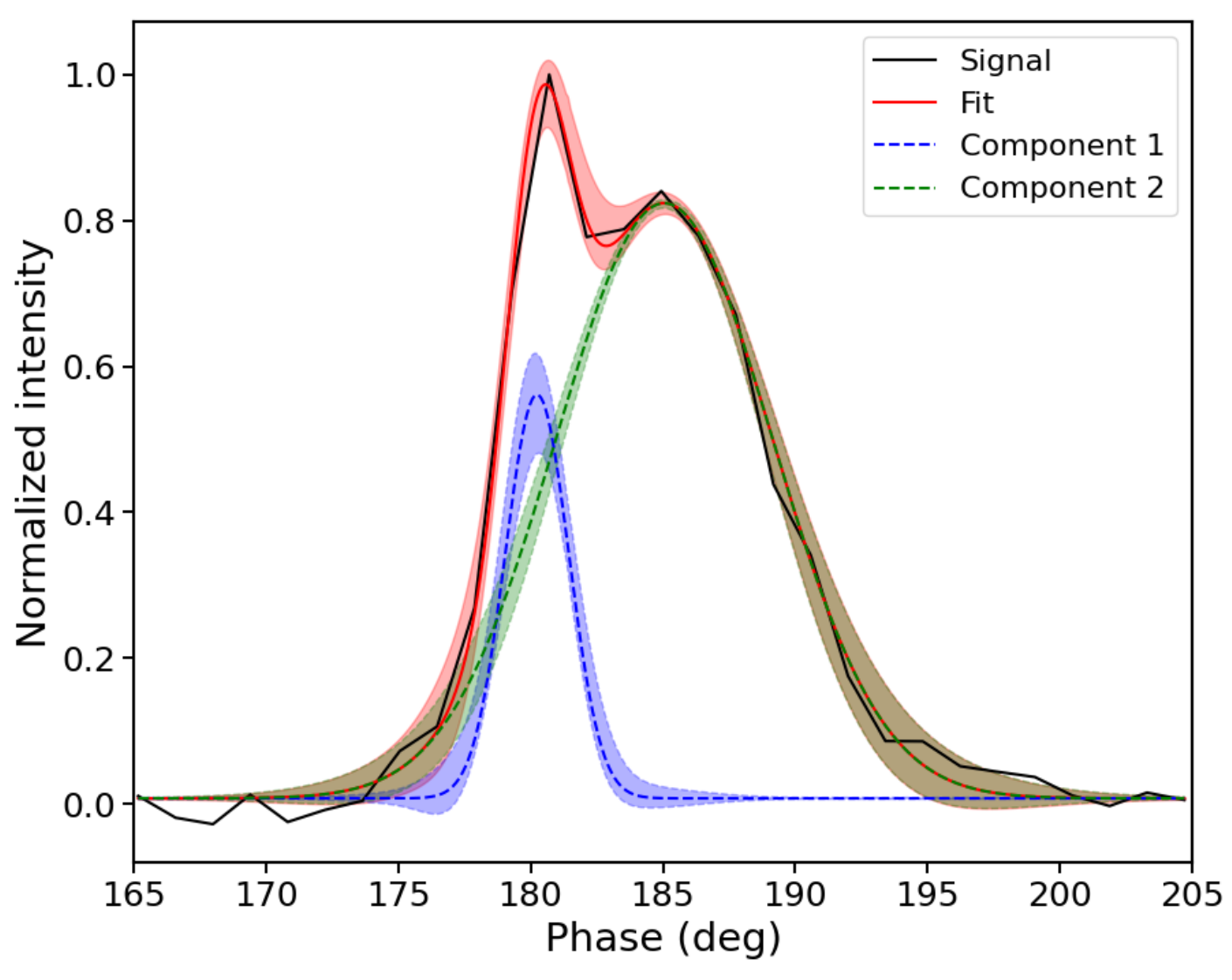}
    \caption{Example of one of the mean profiles fitted with two Gaussian components with their $1\,\sigma$ uncertainties.}
    \label{fig: pulse}
\end{figure}

All the mean profiles look like a double-peak signal as shown in Fig.~\ref{fig: pulse}. We therefore modelled the profiles as the sum of two Gaussian components as:
\begin{equation}\label{eq:gaussian}
    Y = A_1 \exp{\left[-\frac{(x-\mu_1)^2}{2\sigma_1^2} \right]} + A_2 \exp{\left[-\frac{(x-\mu_2)^2}{2\sigma_2^2}\right]} + C ,
\end{equation}
where $A_{1,2}$ are the amplitudes of the leading (left) and trailing (right) components respectively and $\mu_{1,2}$ are the position of their centres. Finally, $\sigma_{1,2}$ are the widths of each component and $C$ is the offset of the signal.

To do the fit, we applied the \textsc{mcmc} method \citep{2021MNRAS.507.2037A} from the \textsc{bilby} library \citep{bilby} to fit Eq.~ \ref{eq:gaussian} to all the mean profiles. We obtained 3 million samples from 10\,000 steps performed with 300 walkers for each fit, with a burn-in of 3333 iterations. An example fit is shown in Fig.~\ref{fig: pulse}. Subsequently, we used 10\,000 posterior samples to obtain the best-fitting parameters of interest with their 1-$\sigma$ uncertainties.

On the one hand, regarding the pulse profile, we studied the evolution of the pulse width at half maximum $W_{50}$, and the shape parameter $S$ (as defined by \cite{2013MNRAS.432.3080K}), which is the ratio between the normalized intensities of the leading peak and the trailing peak.
On the other hand, we studied the evolution of the Gaussian components by characterizing the evolution of the distance between their centers, $\Delta=\mu_2 - \mu_1$, the ratio between their amplitudes, $r_A=A_2/A_1$, and between their widths, $r_W=\sigma_2/\sigma_1$.

%
\section{Results}\label{sec: results}
%

\subsection{Timing solution and glitch characterization}

We assumed the glitch epoch reported by \cite{2022ATel15622....1S} and used a code based on Nested Sampling\footnote{\url{https://github.com/kbarbary/nestle}} \citep{2004AIPC..735..395S} to determine the best timing solution for the glitch parameters. The code employs the PINT software \citep{2021ApJ...911...45L}, with a short data span of $\sim100~\mathrm{d}$ around the glitch in order to not to be affected by red noise.  In addition to the recovery term reported in \cite{2024A&A...689A.191Z}, we found a second and shorter recovery term with a marginal logarithmic Bayes factor of 1.3. We did not find evidence of change in $\ddot \nu$.
In Table~\ref{tab:Vglitch} we present the updated parameters for the timing solution including the glitch event.

\begin{table}
  \centering    
  \caption{Parameters of the timing model for the 2022 glitch in \psr and their 1$\sigma$ uncertainties, obtained by fitting the data spanning from MJD~59683 to MJD~60149.}
   \begin{tabular}{ll}        
     \hline
     Parameter & Value \\
     \hline
     $t_0$ (MJD)& 59000\\
     $\nu(\mathrm{s^{-1}})$ & $5.9960548719(4)$ \\
     $\dot\nu(\mathrm{s^{-2}})$ &  $-6.0348(1)\times 10^{-13}$\\
     $\ddot{\nu}(\mathrm{s^{-3}})$ &  -\\
     $\mathrm{DM}(\mathrm{cm^{-3} pc})$ & 73.728(1) \\
     $t_\mathrm{g}$ (MJD) & 59839.4(5) \\
     $\Delta\nu_\mathrm{p}$ (s$^{-1}$) & $ 2.5619(1)\times 10^{-5}$\\
     $\Delta\dot{\nu}_\mathrm{p}$ (s$^{-2}$)&$-4.1(4)\times 10^{-16}$\\
     $\Delta\ddot{\nu}$\quad (s$^{-3}$)& -\\
     $\Delta\nu_\mathrm{d1}$\, (s$^{-1}$)&$1.23(2)\times 10^{-7}$\\
     $\Delta\nu_\mathrm{d2}$\, (s$^{-1}$)&$8(5)\times 10^{-8}$\\
     $\tau_\mathrm{d1}$ (days) & $48(2)$ \\
     $\tau_\mathrm{d2}$ (days) & $8(5)$\\
     $\Delta \phi$ & $0.14(2)$\\
     $\Delta\nu_\mathrm{g}/\nu$ & $4.306(8)\times 10^{-6}$\\
     $\Delta\dot\nu_\mathrm{g}/\dot\nu$ & 0.24(16)\\
     $Q_1 (\%)$ & 0.476(7) \\
     $Q_2$ & 0.3(2)\\
     $Q$ & 0.8(2) \\
     \hline
   \end{tabular}
  \label{tab:Vglitch}
  \tablefoot{The DM value was taken from ATNF\footnote{\url{https://www.atnf.csiro.au/research/pulsar/psrcat/}}. The glitch epoch and its uncertainty was taken from \citep{2022ATel15622....1S}.}
 \end{table}

\begin{figure}[h]
    \includegraphics[width=\linewidth]{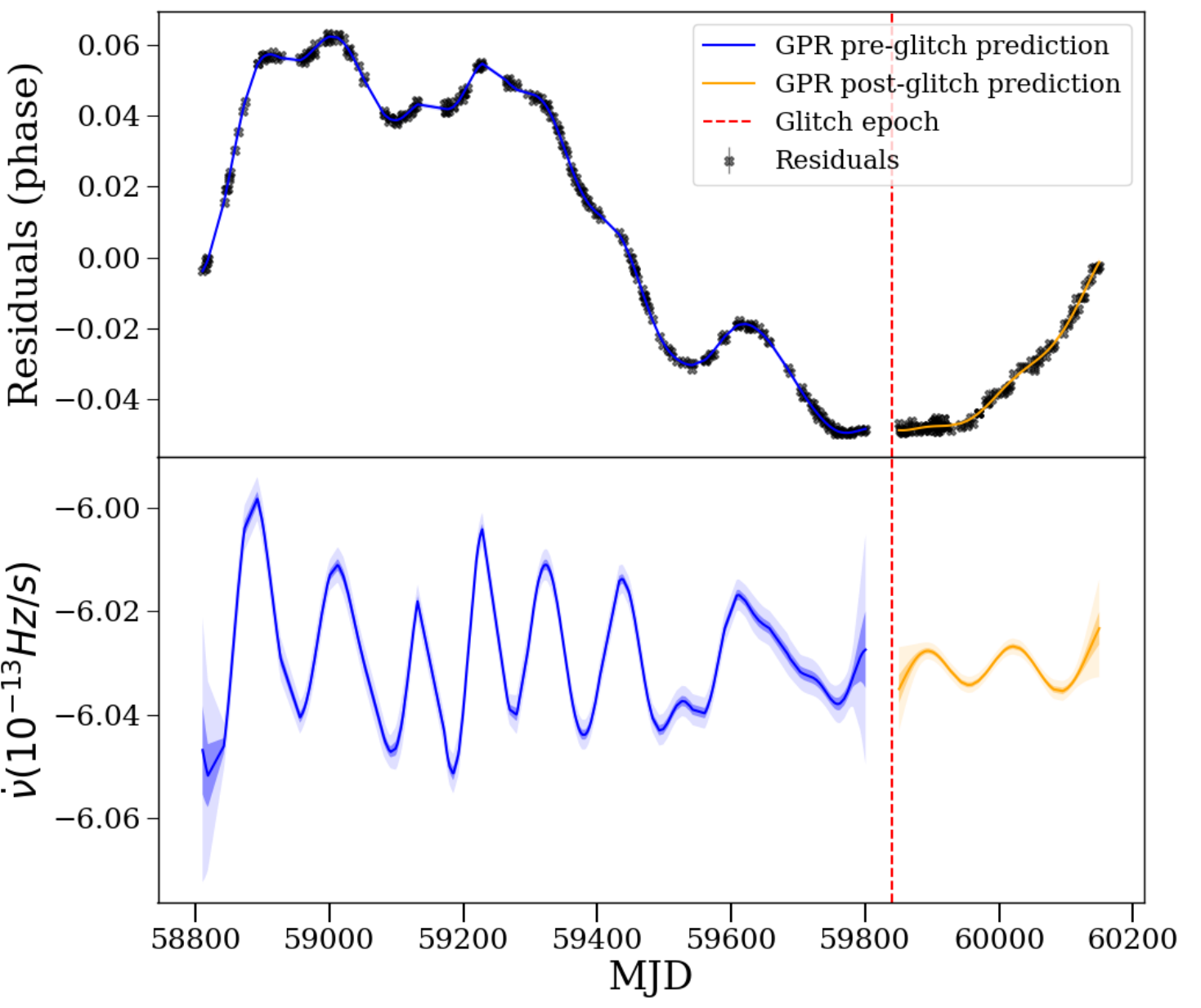}
    \caption{Top panel: residuals after fitting the glitch (black dots) and the GPR prediction (solid curves). Bottom panel: Evolution of $\dot\nu$ obtained from the GPR prediction after the glitch was removed from the residuals; the shaded regions correspond to the 1-$\sigma$ confidence intervals.}
    \label{fig: F1-evolution}
\end{figure}

We used the GPR technique to calculate the oscillation of $\dot\nu$ around the rotational model shown in Table~\ref{tab:Vglitch}, following the same procedure as \cite{2022MNRAS.513.5861S} and \cite{2013MNRAS.432.3080K} to obtain an unbiased solution for the oscillations of $\dot\nu$ in the presence of red noise. We performed independent GPR predictions for the pre-glitch and the post-glitch residuals. In Fig.~\ref{fig: F1-evolution} we show the post-fit timing residuals with the GPR prediction, and the second derivative of the GPR prediction, which is $\dot\nu$. Fig.~\ref{fig: F1-evolution} shows that the standard deviation of $\dot\nu$ is much smaller after the glitch ($\sigma_{\dot\nu} / \dot\nu\sim 0.17\%$) than before the glitch ($\sigma_{\dot\nu} / \dot\nu \sim 0.82\%$).

\begin{figure}[h]
    \includegraphics[width=\linewidth]{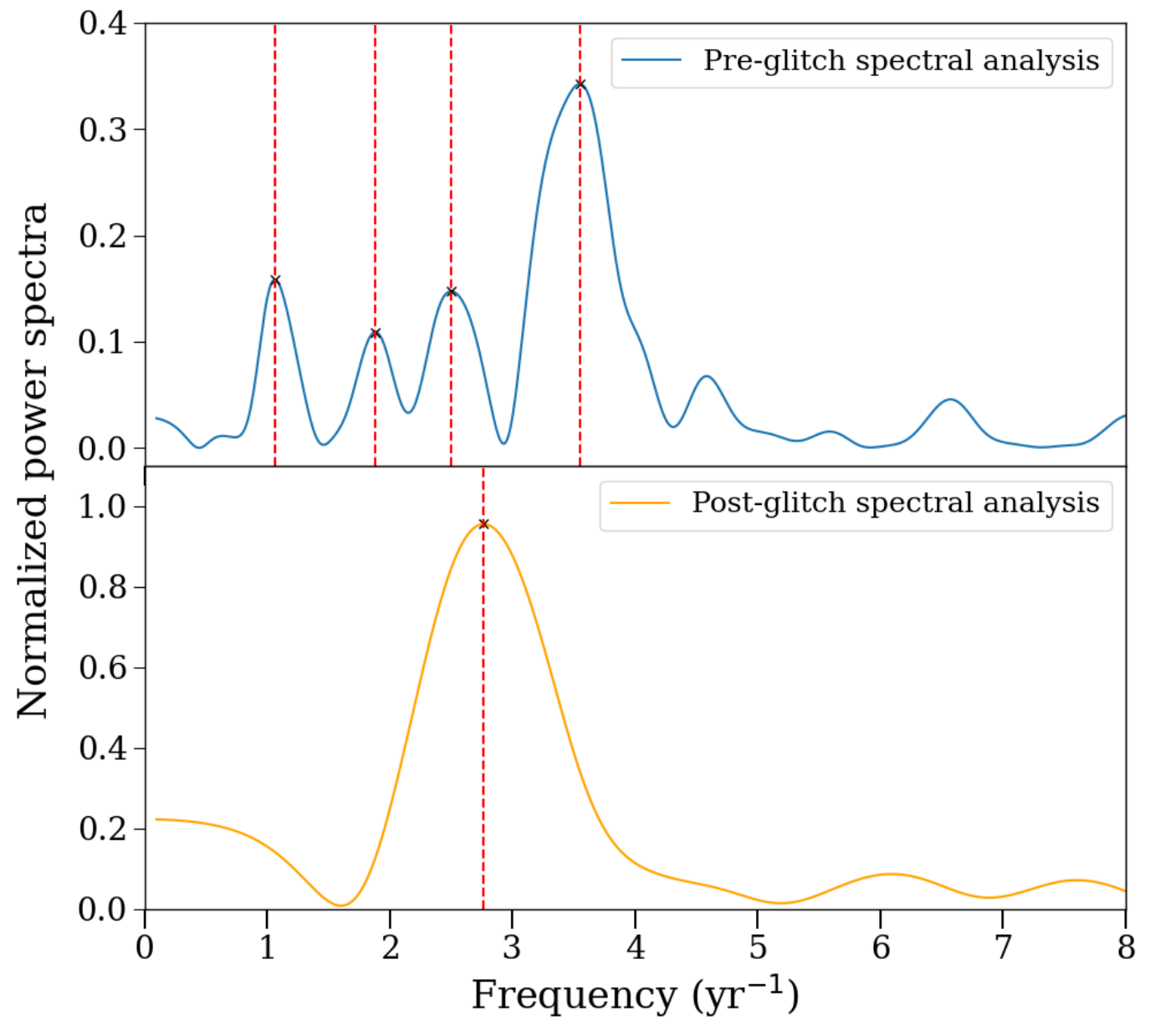}
    \caption{Top panel: Lomb-Scargle analysis of the pre-glitch GPR prediction for $\dot\nu$. Bottom panel: Lomb-Scargle analysis of the post-glitch GPR prediction for $\dot\nu$. All the peaks pointed have a false alarm probability lower than $1\%$.}
    \label{fig: Lomb-Scargle}
\end{figure}

In addition, oscillations seem to have higher-frequency components before the glitch than after the glitch. In order to quantify it, we performed a Lomb-Scargle analysis for the pre-glitch oscillations and the post-glitch oscillations independently. The results are shown in Fig.~\ref{fig: Lomb-Scargle}.
We found four frequency components for the pre-glitch $\dot\nu$ oscillations, while we detected only one for the post-glitch $\dot\nu$ oscillations. The peaks detected in the Lomb-Scargle analysis are enumerated in Table \ref{tab: Lomb}, ordered by their relative intensities in the Lomb-Scargle analysis.

\begin{table}
  \centering    
  \caption{Frequencies of the oscillations of $\dot\nu$ detected with the Lomb-Scargle analysis. We took the FWHM for the uncertainty of the peaks and only considered peaks with false alarm probability lower than $1\%$.}
   \begin{tabular}{lll}        
     \hline
     Peak & Pre-glitch frequency& Post-glitch frequency\\
      & ($\mathrm{yr}^{-1}$) & ($\mathrm{yr}^{-1}$) \\
     \hline
     Primary & 3.5(7) & 2.7(12)\\
     Secondary & 1.1(3) & - \\
     Third & 2.5(5) & - \\
     Fourth & 1.9(3) & - \\
     \hline
   \end{tabular}
  \label{tab: Lomb}
 \end{table}

\subsection{Pulse profile evolution}\label{sec: method}

\begin{figure}[h!]
    \includegraphics[width=\linewidth]{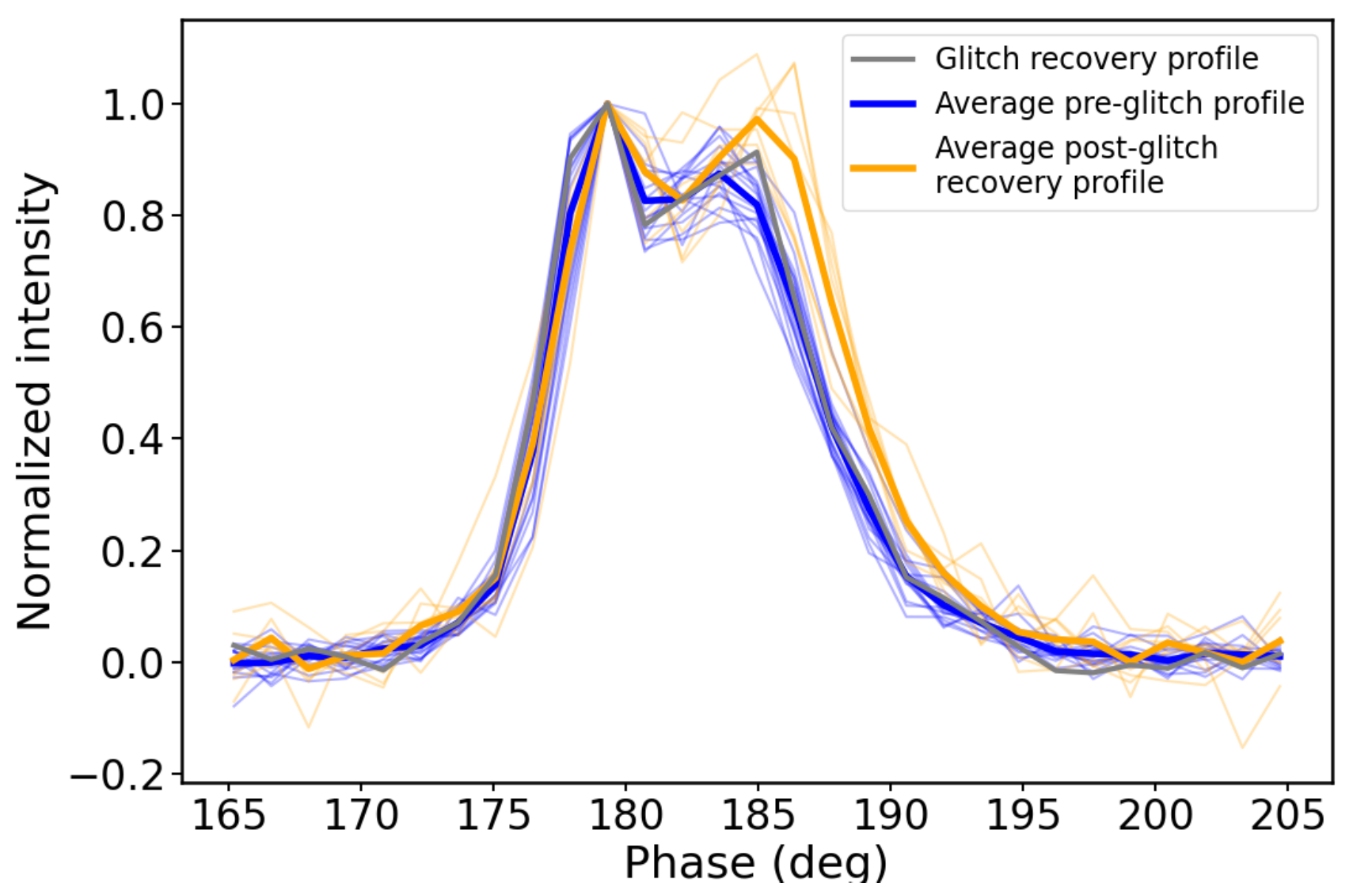}
    \caption{All the normalized mean pulse profiles studied in this work with the pre-glitch and post-glitch recovery mean pulse (solid blue and yellow lines respectively). The profile of the glitch recovery data span is marked with a solid grey line.}
    \label{fig: pulses_all}
\end{figure}

We analyzed 26 mean profiles. 18 of them are pre-glitch observations between MJD 58810 and MJD 59800, and 8 are post-glitch  observations between MJD 59851 and MJD 60149. From the post-glitch profiles, one of them corresponds to a date when the glitch had not finished its recovery phase, while the rest correspond to the post-glitch recovery data span. We show all mean profiles normalized by the amplitude of the leading peak in Fig.~\ref{fig: pulses_all}.

The pulses shown in Fig.~\ref{fig: pulses_all} reveal that for all pre-glitch pulses, the leading peak is greater in amplitude than the trailing peak. However, for some of the post-glitch pulses, the trailing peak is greater than the leading one. Furthermore, for some of the post-glitch pulses, the trailing peak is further away in phase from the leading peak than for the pre-glitch pulses. 
Fig. \ref{fig: pulses_all} also shows the average normalized pulse profile before the glitch and after the glitch recovery (solid blue and yellow lines), clearly exposing the changes in the amplitude of the trailing peak and the distance from the leading peak.
In addition, we marked the only profile that is within the post-glitch recovery data span. This profile looks like a transition between the pre-glitch profile and post-glitch recovery profile.

\begin{figure}[h]
    \includegraphics[width=\linewidth]{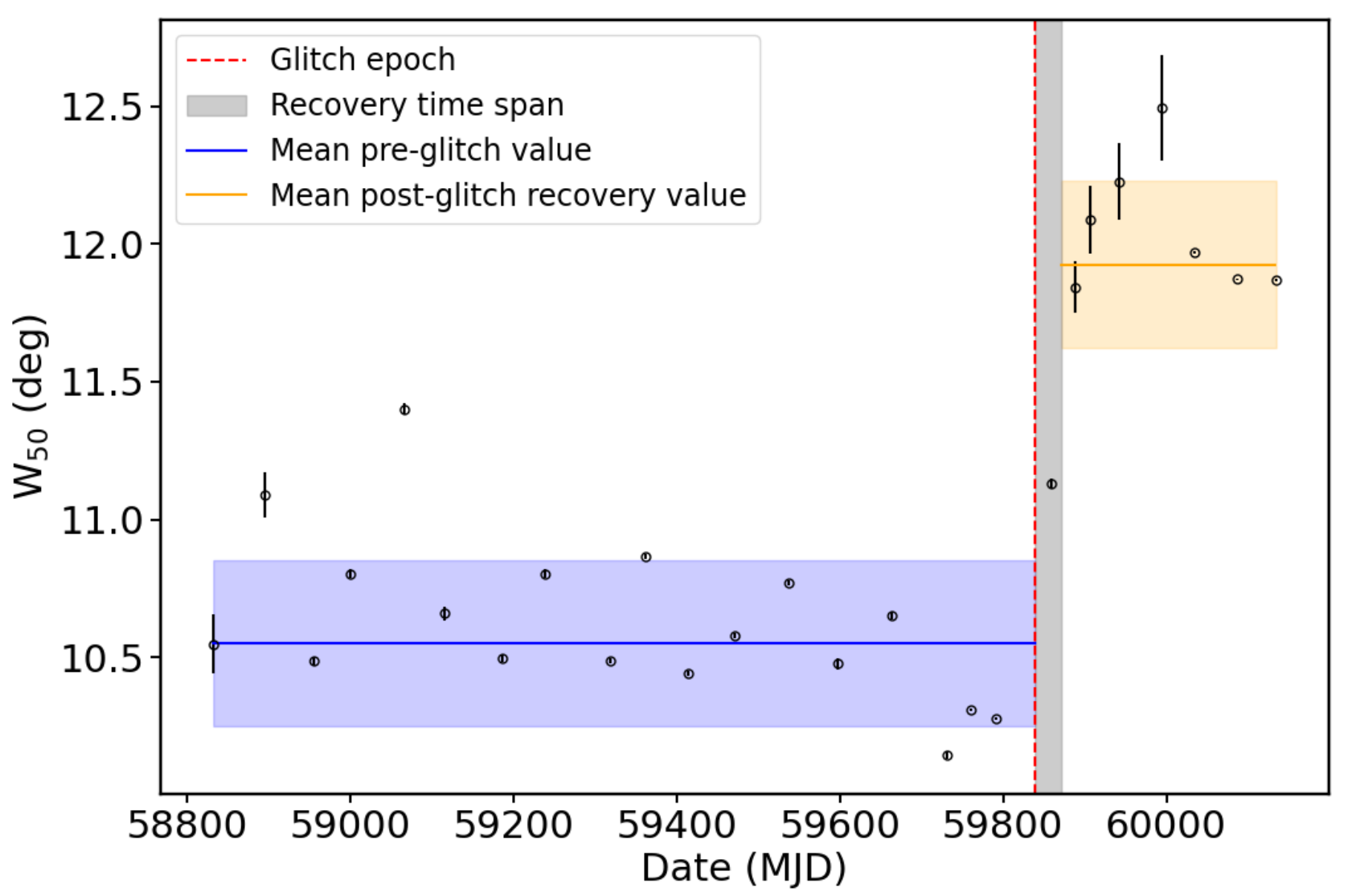}
    \caption{The evolution of $W_{50}$. The horizontal bands correspond to the $1\sigma$ mean values in the pre-glitch (blue) and post-glitch recovery (orange). The vertical grey band marks the glitch recovery timespan.}
    \label{fig: w50}
\end{figure}

We then fitted all the pulses with a two-component Gaussian model, as explained in Sec \ref{sec: gaussian}. From the posteriors, we studied the behaviors of $W_{50}$, $\Delta$, $S$, $\mathrm{r_A}$ and $\mathrm{r_W}$.

\begin{figure}[h!]
    \includegraphics[width=\linewidth]{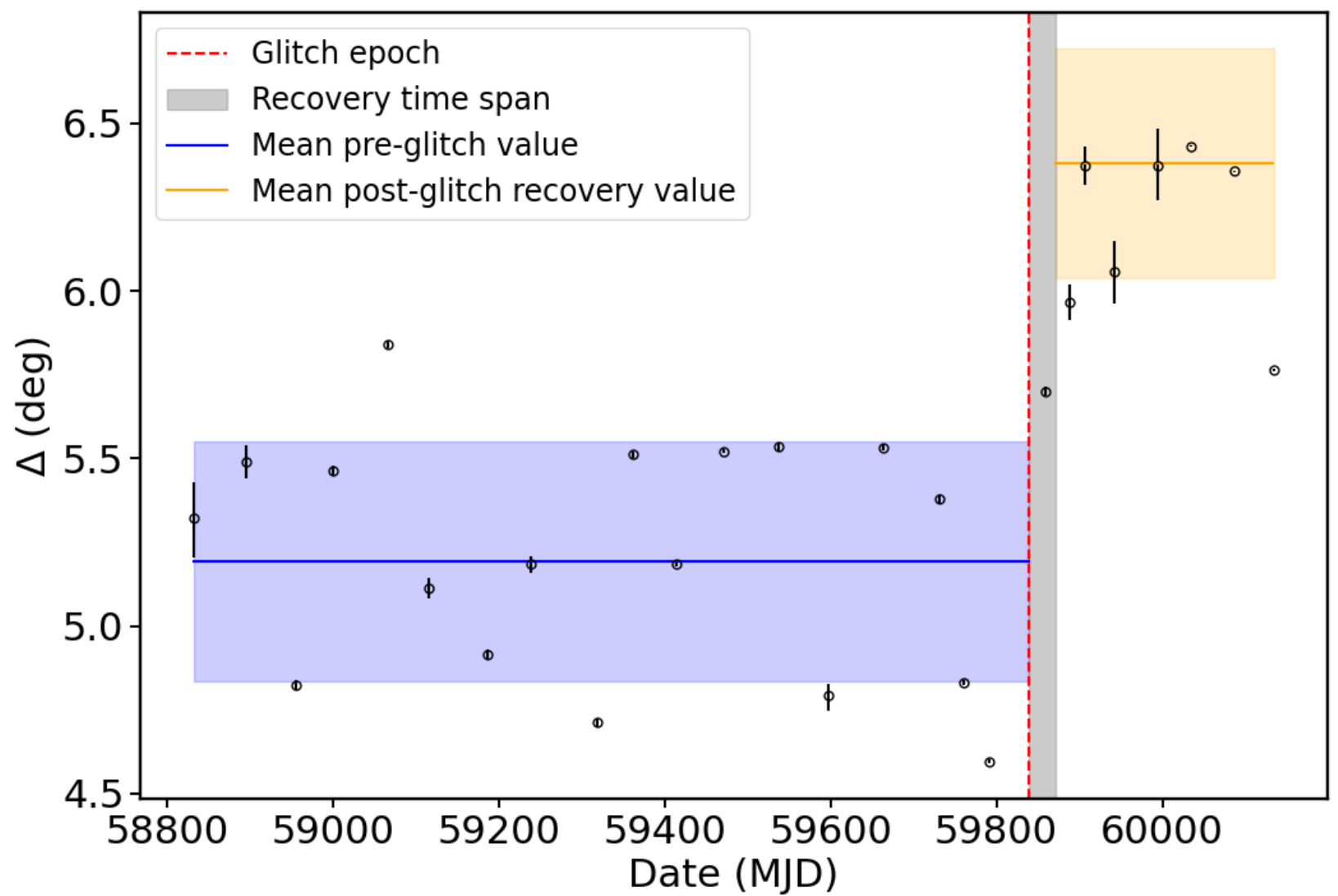}
    \caption{Evolution of the distance between the centers of the Gaussian components.  The horizontal bands corresponds to the $1\sigma$ mean pre-glitch value and the $1\sigma$ mean post-glitch recovery value, respectively. The vertical gray band marks the 33-days glitch recovery epoch.}
    \label{fig: dist_centros}
\end{figure}

In Figs.~\ref{fig: w50} and \ref{fig: dist_centros} we show the evolution of $W_{50}$ and $\Delta$ respectively. These parameters behave similarly between each other as expected. Fig \ref{fig: w50} shows that before the glitch, $W_{50}$ is generally below $11^\circ$, while after the glitch, except for the glitch recovery pulse profile, it is above $11.5^\circ$. The glitch recovery value of $\mathrm{W}_{50}$ may indicate a transition between its pre-glitch and post-glitch value. 

Something similar happens with $\Delta$. Fig. \ref{fig: dist_centros} shows that $\Delta$ lies below $5.6^\circ$ in most of the pre-glitch pulse profiles, and above $5.75^\circ$ for the majority of the post-glitch cases. It also shows an intermediate value of $\Delta$ for the glitch recovery pulse profile. This behavior describes the increasing distance of the trailing peak with respect to the leading peak after the glitch as shown in Fig. \ref{fig: pulses_all}, indicating the Gaussian components are more separated after the glitch.

\begin{figure}[h!]
    \includegraphics[width=\linewidth]{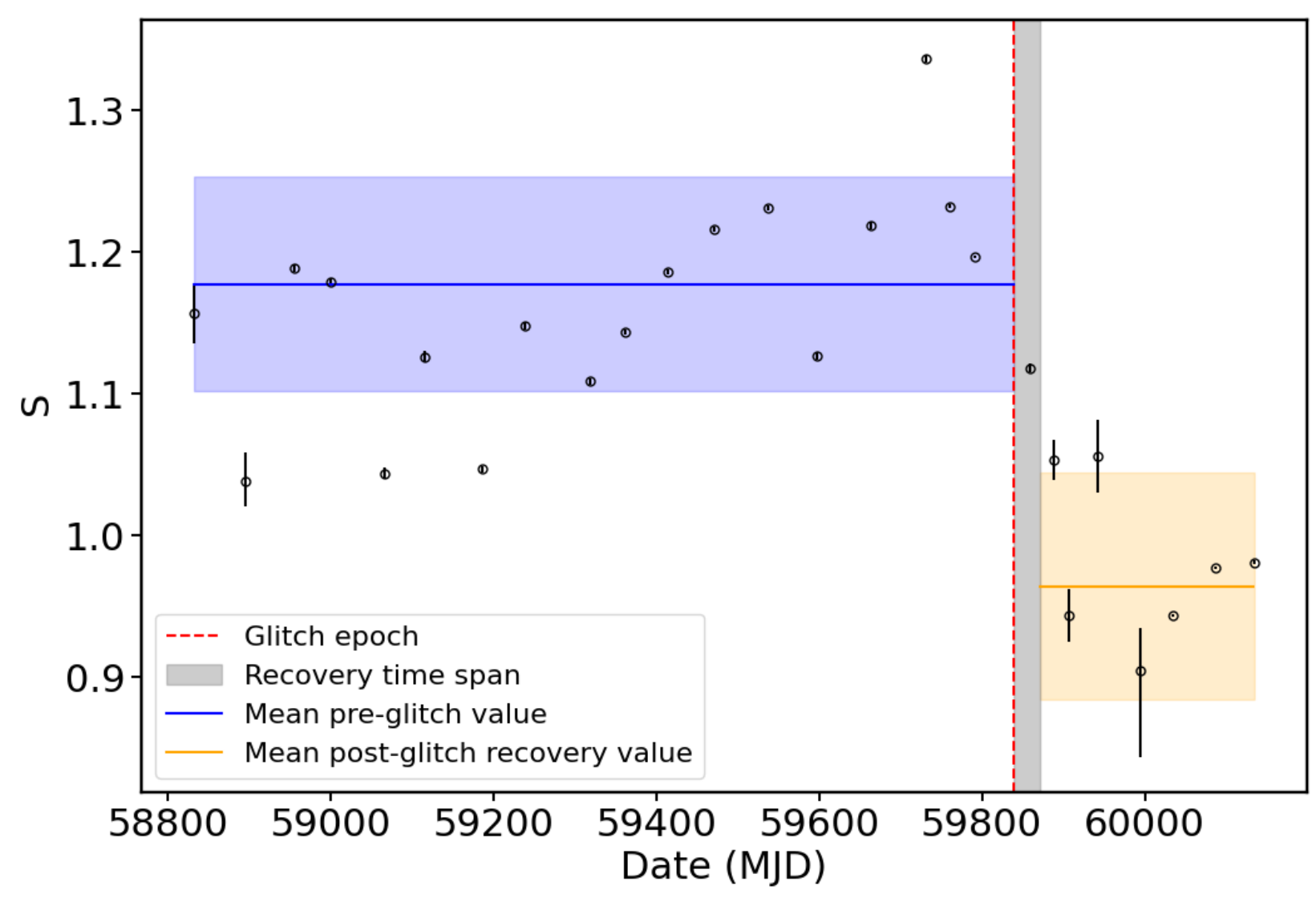}
    \caption{Evolution of the shape parameter $S$.  The horizontal bands correspond to the $1\sigma$ mean pre-glitch value and the $1\sigma$ mean post-glitch recovery value respectively. The vertical grey band marks the glitch recovery data span.}
    \label{fig: S}
\end{figure}

Regarding the shape parameter, $S$, Fig. \ref{fig: S} shows that for all the pulse profiles before the glitch, $S>1$, which means that the leading peak is always greater than the trailing one. However, for the post-glitch recovery profiles, only 2 cases correspond to $S>1$, while the rest of them have clearly $S<1$. Therefore the trailing peak surpassed in amplitude the leading peak after the glitch. Again, the glitch recovery value of $S$ is intermediate between the pre-glitch mean value of $S$ and its post-glitch recovery mean value.

\begin{figure}[h]
    \includegraphics[width=\linewidth]{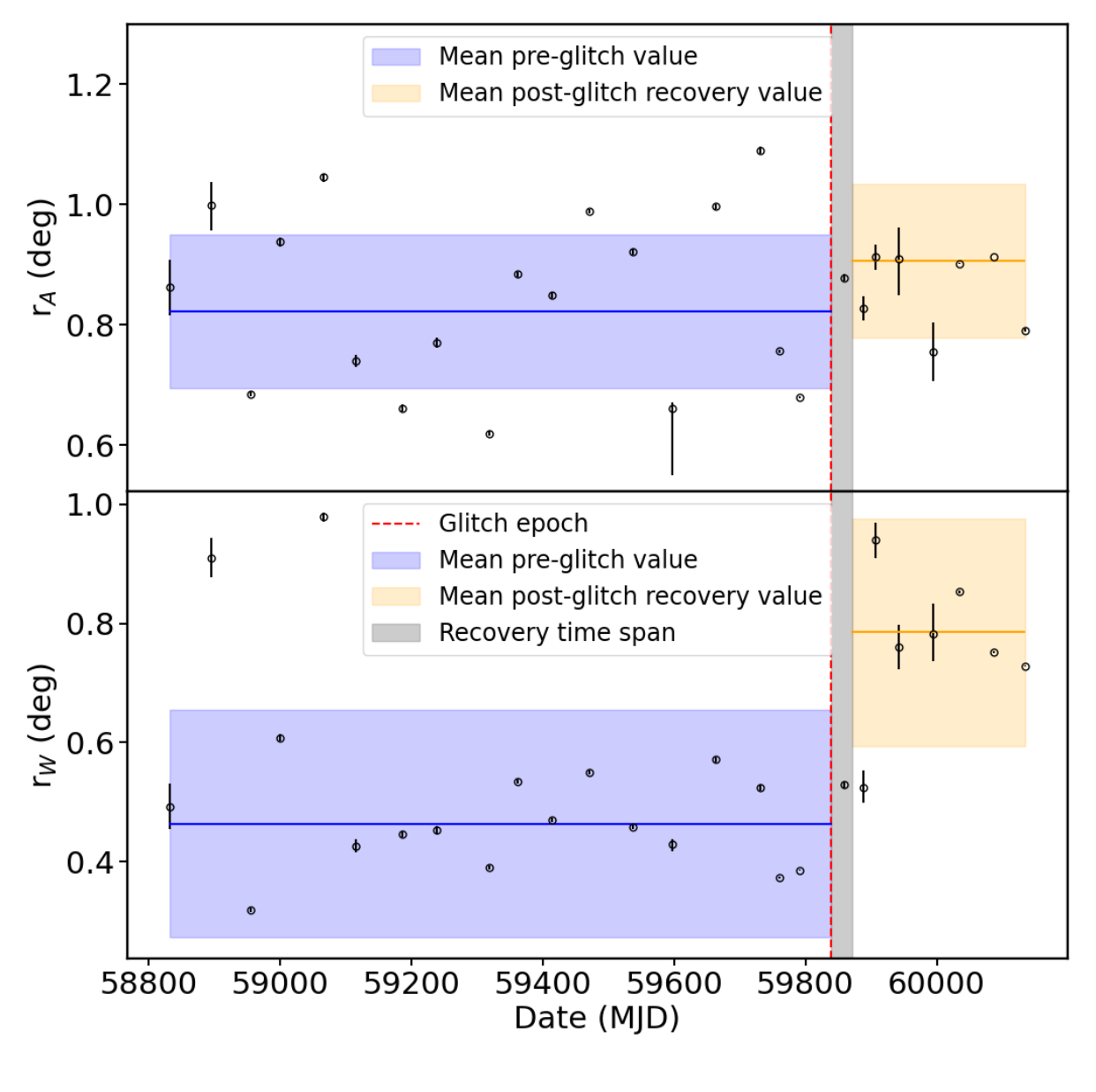}
    \caption{Evolution of the ratio between the amplitudes and the widths of the Gaussian components. The horizontal bands correspond to the $1\sigma$ mean pre-glitch value and the $1\sigma$ mean post-glitch recovery value respectively. The vertical grey band marks the glitch recovery data span.}
    \label{fig: Ramp}
\end{figure}

We also studied the evolution of the Gaussian components by comparing their parameters through $r_{A}$ and $r_{W}$. Fig. \ref{fig: Ramp} shows the evolution of the ratio between the amplitudes of the components, $r_{A}$.
Although there is no significant change in the value of $r_{A}$, the dispersion of the values after the glitch are much smaller than before the glitch, which might be an indicator of the stability of the magnetosphere. With respect to $r_{W}$, as seen in Fig. \ref{fig: Ramp}, the width of the trailing component increases significantly with respect to the leading one, and also presents an intermediate value for the glitch recovery profile.

To quantify the change in the profile after the glitch, in Table~\ref{tab: parameters} we take the $1\sigma$ value of all the parameters studied above. We also included the standard deviation of $\dot\nu$ before and after the glitch. We did not study separated oscillations for the glitch recovery data span given that it is too short to perform adequate GPR predictions. From Table~\ref{tab: parameters} we conclude that $W_{50}$, $\Delta$ and $\mathrm{r_W}$ indicate that the pulse profile is wider after the glitch due to the widening of the trailing Gaussian component. Also, $S$ shows that the peak of the trailing component was amplified after the glitch with respect to the leading component. This may indicate that only the trailing component became brighter after the glitch, which also resulted in a widening of that component. Regarding $r_{A}$, the mean value did not change significantly after the glitch, but the dispersion is $2.5$ times smaller after the glitch than before it. For all the parameters studied here, the values that correspond to the glitch recovery profile lie between their pre-glitch and post-glitch recovery values. We also note that, together with the change in the profile, the oscillations of $\dot\nu$ also changed, decreasing significantly after the glitch.

\begin{table}[htbp]
    \centering
    \caption[]{Comparison of parameters of interest before the glitch, during the glitch recovery, and afterwards. $\sigma_{\dot\nu} / \dot\nu ~(\%)$ refers to the relative standard deviation of $\dot\nu$.}
    \label{tab: parameters}
    \centering
    \begin{tabular}{cccc}
        \hline
        \noalign{\smallskip}
        Parameter & Pre-glitch & Recovery phase & Post-recovery\\
        \noalign{\smallskip}
        \hline
        \noalign{\smallskip}
        $W_{50} ~(\mathrm{deg})$ & 10.6(3) & 11.13(4) & 11.9(3)\\
        $\Delta ~(\mathrm{deg})$ & 5.2(3) & 5.70(3) & 6.4(3) \\
        $\mathrm{S}$ & 1.18(7) & 1.118(7) & 0.96(8) \\
        $r_{A}$ & 0.82(15) & 0.88(1)& 0.90(15)\\
        $r_{W}$ & 0.46(17) & 0.53(11)& 0.79(1889)\\
        $\sigma_{\dot\nu} / \dot\nu ~(\%)$ & 0.82 & - & 0.17\\
        \noalign{\smallskip}
        \hline
    \end{tabular}
    \tablefoot{Best-fitting values and 1-$\sigma$ errors shown here were obtained of the posterior distributions from the MCMC chains.}
\end{table}

%
\section{Discussion}\label{sec:discussion}
%

Generally, glitches are thought to be driven by the superfluid interior of the neutron star, while pulse profile evolution and spin-down oscillating behaviors are considered to be driven by magnetospheric changes \citep{1969Natur.224..673B,2022atcc.book..219A}. However, in some pulsars, there have been detections of changes in the spin-down behavior or pulsar emission after a glitch. In the case of \psr, \cite{2013MNRAS.432.3080K} detected a change in the correlation between the spin-down frequency derivative and the pulse profile shape after its 2009 glitch. \cite{2021RAA....21...42D} and \cite{2022MNRAS.513.5861S} showed that there are different states for the correlation between the spin-down rate but that not all the changes in the spin-down rate are associated with glitches. 

The glitch observed in 2022 in \psr was significantly larger than any previously reported events for this pulsar, with a fractional frequency jump almost 50 times greater than the largest glitch recorded prior to 2022 \citep{2022MNRAS.510.4049B}. This event resulted in a fractional frequency increase of $\frac{\Delta\nu}{\nu}\sim 4300 \times 10^{-9}$ \citep{2022ATel15622....1S,2023MNRAS.521.4504Z}, while previous glitches typically exhibited relative increases of $\frac{\Delta\nu}{\nu} \leq 92 \times 10^{-9}$.

In this work we studied the spin-down rate evolution and the pulse profile behavior across the 2022 giant glitch and we detected a clear change in both of them induced by the glitch. This glitch-induced variation in the pulsar emission thrusts the study of the complex relation between the pulsar rotation and the pulse-profile emission. For example, studying radiative changes during glitches detected "in live" could be a key to understand the magnetospheric structure of neutron stars \citep{2018Natur.556..219P}.

\cite{2022MNRAS.513.5861S} showed that the standard deviation of $\dot\nu$ around the timing model varies with time is not stable, and \cite{2021RAA....21...42D} found that, in some cases, the average value of $\dot\nu$ changed after the pulsar had gone through a glitch. Here, we reported in Table \ref{tab: parameters} the change in $\dot\nu$ due to the glitch. Also, we showed that the standard deviation around $\dot\nu$ is
 $\sigma_{\dot\nu} / \dot\nu \sim 0.82\%$ for most of the 1000 days in our pre-glitch data span, while for the 300 days of observations after the glitch, the standard deviation decreased significantly down to $\sigma_{\dot\nu} / \dot\nu \sim 0.17\%$ (bottom panel of Figure~\ref{fig: F1-evolution}). Considering the scenario where the glitch is triggered when the critical lag between the superfluid interior of the star and the rigid crust is reached \citep{2022RPPh...85l6901A,2016MNRAS.461L..77H}, the post-glitch decrease in the standard deviation of $\dot\nu$ may indicate that the system rapidly reestablishes equilibrium as the superfluid and crust regain rotational synchronization.
 
We also found a significant change in the periodicity spectra of the spin-down rate evolution (Fig.~\ref{fig: Lomb-Scargle}). In \cite{2022MNRAS.513.5861S}, they found the periodicity of the $\dot \nu$ evolution had a peak in $2.8(1)~\mathrm{yr}^{-1}$, although the spectra is highly noisy and with many peaks. Here, although with a much smaller data span, we find a clearer spectra with four peaks before the glitch, and only one peak after the glitch. Both of these results are consistent with our result in \cite{2024A&A...689A.191Z}, where we concluded that red noise had diminished after the glitch. Considering that we detect four frequencies contributing to red noise before the glitch, and only one after the glitch, such red-noise components may correspond to different regions inside the neutron star responsible for the pulsar braking torque. Assuming that the oscillations are driven by the coupling between different regions of the superfluid interior and the solid crust \citep{2023MNRAS.518.5734G,2017MNRAS.471.4827G}, these regions might decouple from the braking torque of the star during the glitch and then re-couple again on different time scales. In addition, if oscillations of the frequency derivative were driven by oscillations in the magnetic field \citep{2012ApJ...761..102Z}, this result might point to a reconfiguration of the magnetosphere during the glitch. The idea of a reconfiguration or a stabilization of the magnetosphere after the glitch is also supported by the fact that the dispersion of $\mathrm{r_A}$ is much smaller after the glitch than before it, together with the more stable behavior of $\dot\nu$. Therefore the evidence gathered here shows that the pulsar magnetosphere was likely affected by the decoupling between the superfluid interior and the crust of the neutron star during the glitch \citep{2018Natur.556..219P}, or by the mechanism that triggered the glitch \citep{2020MNRAS.496.2506G}. 

In addition, \cite{2024arXiv240809204Z} also detected that the typical periodicity and amplitude of $\dot\nu$ oscillation in PSR J1522$-$5735 changed after each glitch, which is a similar behavior that we detected for PSR J0742$-$2822. This was interpreted in terms of the vortex creep model. During external driven events like glitches, the vortices that are pinned to the neutron star's crust are unpinned and then repinned in new positions. This rearrangement alters the vortices oscillations around their equilibrium configuration. As a result, the frequency and amplitude of these oscillations change. These vortex oscillations generate variable torques on the crust, leading to fluctuations in the pulsar's spin-down rate ($\dot{\nu}$) after each glitch \citep{2023MNRAS.518.5734G}.

We also detected that the transition between the pre-glitch and post-glitch magnetospheric states is not discrete. We found evidence for an intermediate magnetospheric state probably related to the glitch recovery epoch. This novel detection opens the door to further study the correlation between the superfluid interior of the neutron star and the different magnetospheric states, suggesting that the latter may be driven by the dynamics of the superfluid or vice-versa, as well as a strong coupling between the magnetosphere and the superfluid dynamics. It is intriguing whether the recovery time scales of the superfluid may also affect the pulsar emission. This could be possible if the magnetosphere is also coupled to the different regions of the superfluid interior with their respective recovery time-scales \citep{2023MNRAS.518.5734G}.

The change in $r_\mathrm{w}$ after the glitch indicates that the widening of the pulse profile is mainly caused by the widening of one of its components, in particular, the trailing peak. Consequently, it may be that the different components of the pulse profile have different coupling strengths to the different magnetic field lines in the magnetosphere and in turn to the superfluid interior of the neutron star, and therefore they are differently affected by the glitch. For example, if the glitch is triggered by a broken plate \citep{2015MNRAS.449..933A,2019MNRAS.488.2275G,2024MNRAS.533.4274L}, this broken plate could be affecting only the region of the polar cap where the trailing peak is emitted.

If the widening of the pulse is only due to a change of the inclination angle $\alpha$ between the magnetic and rotational axes, \cite{1990ApJ...352..247R} calculated that $W_{50}$ is linked to them as:
\begin{equation}\label{eq: w50}
   W_{50} = 2.45^\circ \frac{P^{-1/2}}{\mathrm{sin}\alpha}.
\end{equation}
%
According to this, the change in $W_{50}$ of $1.3(5)^\circ$ during the glitch corresponds to $\Delta \alpha \sim 4.8^\circ$. This value is quite high compared to the $\Delta \alpha \sim 0.08^\circ$ inferred for PSR J1048$-$5832 by \cite{2024MNRAS.533.4274L}. Given that the $\Delta \alpha$ value is too large, we conclude that the widening of the pulse is not solely linked to a change in the inclination angle but also to pure magnetospheric changes. 

We can confirm that the $\Delta \alpha$ obtained before is invalid using the equations for the vortex creep model presented in \cite{2024MNRAS.533.4274L}. The persistent shift in the spin-down rate is (Eq.~\ref{eq: spin-down}):
\begin{equation}\label{eq: spin-down}
    \frac{\Delta\dot\nu}{\dot\nu}_\mathrm{per,total}=\frac{\Delta\dot\nu}{\dot\nu}_\mathrm{per,shift} + \frac{\Delta\dot\nu}{\dot\nu}_\mathrm{per,trap},
\end{equation}
where $\frac{\Delta\dot\nu}{\dot\nu}_\mathrm{per,trap}$ is the persistent step in the spin-down rate given by the formation of new vortex traps and $\frac{\Delta\dot\nu}{\dot\nu}_\mathrm{per,shift}$ is the contribution to the persistent change in the spin-down rate given by the change in the inclination angle between the magnetic axis and the rotational axis. Using Eq.~\ref{eq: spin-down} and that $\frac{\Delta\dot\nu}{\dot\nu}_\mathrm{per,trap}\geq0$, we obtain:
\begin{equation}\label{eq: spin-down_shift}
    \frac{\Delta\dot\nu}{\dot\nu}_\mathrm{per,total}\geq\frac{\Delta\dot\nu}{\dot\nu}_\mathrm{per,shift},
\end{equation}
where \citep{2017ApJ...842...53Z}:

\begin{equation}\label{eq: shift}
    \frac{\Delta\dot\nu}{\dot\nu}_\mathrm{per,shift}=\frac{\sin(2\alpha)\Delta\alpha}{1+\sin^2(\alpha)}.
\end{equation}

From Tab. \ref{tab:Vglitch}, we obtain that the persistent change in the spin-down rate is $\frac{\Delta\dot\nu}{\dot\nu}_\mathrm{per,total} \sim 7\times10^{-4}$. Therefore, $\Delta \alpha \leq 0.05^\circ$, which is a value similar to the the one obtained by \cite{2024MNRAS.533.4274L}. This way we conclude that the widening of the pulse is not only because of a change in $\alpha$.

In summary, our results highlight the complexity of the interaction between the internal dynamics of the superfluid and the magnetospheric evolution in \psr \citep{2010Sci...329..408L,2013MNRAS.432.3080K,2021RAA....21...42D,2022MNRAS.513.5861S}. We demonstrated that the 2022 giant glitch clearly affected the magnetosphere of the pulsar and, consequently, both the oscillations in $\dot\nu$ and the variation in the pulse profile were significantly altered. The reduction in red noise \citep{2024A&A...689A.191Z} and the decrease in the standard deviation of $\dot\nu$ after the 2022 glitch suggest that the system stabilized after the glitch. In addition, the detection of a transient profile during the recovery period after the glitch raised may provide insight into the underlying mechanisms of glitches and their associated recovery processes.

Finally, we emphasize the need for long-term continued monitoring of glitching pulsars such as \psr to better understand the link between glitches and magnetospheric changes. Given that this was the first time that an abrupt change in the pulse profile of \psr was detected after the glitch, and that this was the largest glitch in this pulsar so far, we speculate that the size of the changes in the magnetosphere are proportional to the size of the glitch. 

%

%
\section{Conclusions} \label{sec:conclusions}
%
In this work we studied the rotational and pulse profile evolution of \psr between MJD 58810 and MJD 60149, including the 2022 giant glitch on MJD 59839, based on high-cadence observations of the pulsar performed with the IAR radio telescopes. We found that both the rotational behavior and the emission profile of the pulsar significantly changed after the glitch.

The oscillations of $\dot \nu$ around the simple rotational model of the pulsar and the pulse profile emission are phenomena thought to be driven by the pulsar magnetosphere. Considering that we found they change after the glitch, it follows that either the magnetosphere is sensitive to changes in the superfluid interior of the star, or that the superfluid plays a more direct role in the behavior of the frequency derivative, and the pulsar emission mechanism.

Furthermore, we also found a state transition in the magnetosphere coincident with the recovery time scale of the glitch, reinforcing the premise of a coupling between the magnetosphere and the superfluid interior of the star. Such conjecture is also strengthened by the fact that, before the glitch, oscillations in the frequency derivative occur at four different timescales, while after the giant glitch the oscillations damp significantly, experiencing a single quasi-periodicity, with a much lower amplitude. This suggests that multiple regions inside the neutron star are responsible for the braking torque. Such regions decouple during the glitch and begin to re-couple again at different timescales.

Future real-time glitch observations, combined with detailed analyses of pulse profile and rotational dynamics, could provide more insights into the relation between pulse-profile evolution and internal processes in neutron stars.


\begin{acknowledgements}
      FG and JAC are CONICET researchers and acknowledge support by PIP 0113 (CONICET). FG acknowledges support by PIBAA 1275 (CONICET). COL gratefully acknowledges the National Science Foundation (NSF) for financial support from Grant No. PHY-2207920. CME acknowledges support from the grant ANID FONDECYT 1211964. EG is supported by National Natural Science Foundation of China (NSFC) programme 11988101 under the foreign talents grant QN2023061004L. We also extend our gratitude to the technical staff at the IAR for their continuous efforts that enable our high-cadence observational campaign. 
      JAC was also supported by grant PID2022-136828NB-C42 funded by the Spanish MCIN/AEI/ 10.13039/501100011033 and “ERDF A way of making Europe” and by Consejería de Economía, Innovación, Ciencia y Empleo of Junta de Andalucía as research group FQM-322.

\end{acknowledgements}

%
\bibliographystyle{aa} 
\bibliography{biblio} 

\begin{thebibliography}{55}
\expandafter\ifx\csname natexlab\endcsname\relax\def\natexlab#1{#1}\fi

\bibitem[{{Akbal} {et~al.}(2015){Akbal}, {G{\"u}gercino{\u{g}}lu},
  {{\c{S}}a{\c{s}}maz Mu{\c{s}}}, \& {Alpar}}]{2015MNRAS.449..933A}
{Akbal}, O., {G{\"u}gercino{\u{g}}lu}, E., {{\c{S}}a{\c{s}}maz Mu{\c{s}}}, S.,
  \& {Alpar}, M.~A. 2015, \mnras, 449, 933

\bibitem[{{Antonelli} {et~al.}(2022){Antonelli}, {Montoli}, \&
  {Pizzochero}}]{2022atcc.book..219A}
{Antonelli}, M., {Montoli}, A., \& {Pizzochero}, P.~M. 2022, in Astrophysics in
  the XXI Century with Compact Stars. Edited by C.A.Z. Vasconcellos. eISBN
  978-981-12-2094-4. Singapore: World Scientific, ed. C.~A.~Z. {Vasconcellos},
  219--281

\bibitem[{{Antonopoulou} {et~al.}(2022){Antonopoulou}, {Haskell}, \&
  {Espinoza}}]{2022RPPh...85l6901A}
{Antonopoulou}, D., {Haskell}, B., \& {Espinoza}, C.~M. 2022, Reports on
  Progress in Physics, 85, 126901

\bibitem[{{Ashton} {et~al.}(2019){Ashton}, {H{\"u}bner}, {Lasky}, {Talbot},
  {Ackley}, {Biscoveanu}, {Chu}, {Divakarla}, {Easter}, {Goncharov}, {Hernandez
  Vivanco}, {Harms}, {Lower}, {Meadors}, {Melchor}, {Payne}, {Pitkin},
  {Powell}, {Sarin}, {Smith}, \& {Thrane}}]{bilby}
{Ashton}, G., {H{\"u}bner}, M., {Lasky}, P.~D., {et~al.} 2019, \apjs, 241, 27

\bibitem[{{Ashton} \& {Talbot}(2021)}]{2021MNRAS.507.2037A}
{Ashton}, G. \& {Talbot}, C. 2021, \mnras, 507, 2037

\bibitem[{{Astropy Collaboration} {et~al.}(2013){Astropy Collaboration},
  {Robitaille}, {Tollerud}, {Greenfield}, {Droettboom}, {Bray}, {Aldcroft},
  {Davis}, {Ginsburg}, {Price-Whelan}, {Kerzendorf}, {Conley}, {Crighton},
  {Barbary}, {Muna}, {Ferguson}, {Grollier}, {Parikh}, {Nair}, {Unther},
  {Deil}, {Woillez}, {Conseil}, {Kramer}, {Turner}, {Singer}, {Fox}, {Weaver},
  {Zabalza}, {Edwards}, {Azalee Bostroem}, {Burke}, {Casey}, {Crawford},
  {Dencheva}, {Ely}, {Jenness}, {Labrie}, {Lim}, {Pierfederici}, {Pontzen},
  {Ptak}, {Refsdal}, {Servillat}, \& {Streicher}}]{2013A&A...558A..33A}
{Astropy Collaboration}, {Robitaille}, T.~P., {Tollerud}, E.~J., {et~al.} 2013,
  \aap, 558, A33

\bibitem[{{Basu} {et~al.}(2022){Basu}, {Shaw}, {Antonopoulou}, {Keith}, {Lyne},
  {Mickaliger}, {Stappers}, {Weltevrede}, \& {Jordan}}]{2022MNRAS.510.4049B}
{Basu}, A., {Shaw}, B., {Antonopoulou}, D., {et~al.} 2022, \mnras, 510, 4049

\bibitem[{{Baym} {et~al.}(1969){Baym}, {Pethick}, \&
  {Pines}}]{1969Natur.224..673B}
{Baym}, G., {Pethick}, C., \& {Pines}, D. 1969, \nat, 224, 673

\bibitem[{{Chilenski} {et~al.}(2015){Chilenski}, {Greenwald}, {Marzouk},
  {Howard}, {White}, {Rice}, \& {Walk}}]{2015NucFu..55b3012C}
{Chilenski}, M.~A., {Greenwald}, M., {Marzouk}, Y., {et~al.} 2015, Nuclear
  Fusion, 55, 023012

\bibitem[{{Dang} {et~al.}(2021){Dang}, {Wang}, {Wang}, {Yuan}, {Shang}, {Yuen},
  {Ge}, {Zhou}, {Wang}, {Kou}, {Yan}, {Wang}, {Wen}, {Bai}, {Liu}, \&
  {Zhou}}]{2021RAA....21...42D}
{Dang}, S.-J., {Wang}, N., {Wang}, H.-H., {et~al.} 2021, Research in Astronomy
  and Astrophysics, 21, 042

\bibitem[{{Espinoza} {et~al.}(2011){Espinoza}, {Lyne}, {Stappers}, \&
  {Kramer}}]{2011MNRAS.414.1679E}
{Espinoza}, C.~M., {Lyne}, A.~G., {Stappers}, B.~W., \& {Kramer}, M. 2011,
  \mnras, 414, 1679

\bibitem[{{Fuentes} {et~al.}(2017){Fuentes}, {Espinoza}, {Reisenegger}, {Shaw},
  {Stappers}, \& {Lyne}}]{2017A&A...608A.131F}
{Fuentes}, J.~R., {Espinoza}, C.~M., {Reisenegger}, A., {et~al.} 2017, \aap,
  608, A131

\bibitem[{{Gancio} {et~al.}(2020){Gancio}, {Lousto}, {Combi}, {del Palacio},
  {L{\'o}pez Armengol}, {Combi}, {Garc{\'\i}a}, {Kornecki}, {M{\"u}ller},
  {Guti{\'e}rrez}, {Hauscarriaga}, \& {Mancuso}}]{2020A&A...633A..84G}
{Gancio}, G., {Lousto}, C.~O., {Combi}, L., {et~al.} 2020, \aap, 633, A84

\bibitem[{{Grover} {et~al.}(2024){Grover}, {Joshi}, {Singha},
  {G{\"u}gercino{\u{g}}lu}, {Arumugam}, {Bandyopadhyay}, {Chibueze}, {Desai},
  {Eya}, {Kundu}, \& {Urama}}]{2024arXiv240514351G}
{Grover}, H., {Joshi}, B.~C., {Singha}, J., {et~al.} 2024, arXiv e-prints,
  arXiv:2405.14351

\bibitem[{{G{\"u}gercino{\u{g}}lu} \& {Alpar}(2020)}]{2020MNRAS.496.2506G}
{G{\"u}gercino{\u{g}}lu}, E. \& {Alpar}, M.~A. 2020, \mnras, 496, 2506

\bibitem[{{G{\"u}gercino{\u{g}}lu} {et~al.}(2022){G{\"u}gercino{\u{g}}lu},
  {Ge}, {Yuan}, \& {Zhou}}]{2022MNRAS.511..425G}
{G{\"u}gercino{\u{g}}lu}, E., {Ge}, M.~Y., {Yuan}, J.~P., \& {Zhou}, S.~Q.
  2022, \mnras, 511, 425

\bibitem[{{G{\"u}gercino{\u{g}}lu} {et~al.}(2023){G{\"u}gercino{\u{g}}lu},
  {K{\"o}ksal}, \& {G{\"u}ver}}]{2023MNRAS.518.5734G}
{G{\"u}gercino{\u{g}}lu}, E., {K{\"o}ksal}, E., \& {G{\"u}ver}, T. 2023,
  \mnras, 518, 5734

\bibitem[{{G{\"u}gercino{\v{g}}lu} \& {Alpar}(2017)}]{2017MNRAS.471.4827G}
{G{\"u}gercino{\v{g}}lu}, E. \& {Alpar}, M.~A. 2017, \mnras, 471, 4827

\bibitem[{{G{\"u}gercino{\v{g}}lu} \& {Alpar}(2019)}]{2019MNRAS.488.2275G}
{G{\"u}gercino{\v{g}}lu}, E. \& {Alpar}, M.~A. 2019, \mnras, 488, 2275

\bibitem[{{Haskell}(2016)}]{2016MNRAS.461L..77H}
{Haskell}, B. 2016, \mnras, 461, L77

\bibitem[{{Haskell} \& {Melatos}(2015)}]{2015IJMPD..2430008H}
{Haskell}, B. \& {Melatos}, A. 2015, International Journal of Modern Physics D,
  24, 1530008

\bibitem[{{Hobbs} {et~al.}(2012){Hobbs}, {Coles}, {Manchester}, {Keith},
  {Shannon}, {Chen}, {Bailes}, {Bhat}, {Burke-Spolaor}, {Champion},
  {Chaudhary}, {Hotan}, {Khoo}, {Kocz}, {Levin}, {Oslowski}, {Preisig}, {Ravi},
  {Reynolds}, {Sarkissian}, {van Straten}, {Verbiest}, {Yardley}, \&
  {You}}]{2012MNRAS.427.2780H}
{Hobbs}, G., {Coles}, W., {Manchester}, R.~N., {et~al.} 2012, \mnras, 427, 2780

\bibitem[{{Hobbs} {et~al.}(2006){Hobbs}, {Edwards}, \&
  {Manchester}}]{2006MNRAS.369..655H}
{Hobbs}, G.~B., {Edwards}, R.~T., \& {Manchester}, R.~N. 2006, \mnras, 369, 655

\bibitem[{{Hotan} {et~al.}(2004){Hotan}, {van Straten}, \&
  {Manchester}}]{2004PASA...21..302H}
{Hotan}, A.~W., {van Straten}, W., \& {Manchester}, R.~N. 2004, \pasa, 21, 302

\bibitem[{{Keith} {et~al.}(2013){Keith}, {Shannon}, \&
  {Johnston}}]{2013MNRAS.432.3080K}
{Keith}, M.~J., {Shannon}, R.~M., \& {Johnston}, S. 2013, \mnras, 432, 3080

\bibitem[{{Kou} {et~al.}(2018){Kou}, {Yuan}, {Wang}, {Yan}, \&
  {Dang}}]{2018MNRAS.478L..24K}
{Kou}, F.~F., {Yuan}, J.~P., {Wang}, N., {Yan}, W.~M., \& {Dang}, S.~J. 2018,
  \mnras, 478, L24

\bibitem[{{Liu} {et~al.}(2022){Liu}, {Wang}, {Shen}, {Yan}, {Tong}, {Huang}, \&
  {Zhao}}]{2022ApJ...931..103L}
{Liu}, J., {Wang}, H.-G., {Shen}, Z.-Q., {et~al.} 2022, \apj, 931, 103

\bibitem[{{Liu} {et~al.}(2024){Liu}, {Yuan}, {Ge}, {Ye}, {Zhou}, {Dang},
  {Zhou}, {G{\"u}gercino{\u{g}}lu}, {Wang}, {Wang}, {Li}, {Li}, \&
  {Wang}}]{2024MNRAS.533.4274L}
{Liu}, P., {Yuan}, J.~P., {Ge}, M.~Y., {et~al.} 2024, \mnras, 533, 4274

\bibitem[{{Lousto} {et~al.}(2024){Lousto}, {Missel}, {Zubieta}, {del Palacio},
  {Garc{\'\i}a}, {Gancio}, {Wang}, {Araujo Furlan}, \&
  {Combi}}]{2024RMxAC..56..134L}
{Lousto}, C.~O., {Missel}, R., {Zubieta}, E., {et~al.} 2024, in Revista
  Mexicana de Astronomia y Astrofisica Conference Series, Vol.~56, Revista
  Mexicana de Astronomia y Astrofisica Conference Series, 134--144

\bibitem[{{Luo} {et~al.}(2021){Luo}, {Ransom}, {Demorest}, {Ray}, {Archibald},
  {Kerr}, {Jennings}, {Bachetti}, {van Haasteren}, {Champagne}, {Colen},
  {Phillips}, {Zimmerman}, {Stovall}, {Lam}, \& {Jenet}}]{2021ApJ...911...45L}
{Luo}, J., {Ransom}, S., {Demorest}, P., {et~al.} 2021, \apj, 911, 45

\bibitem[{{Lyne} {et~al.}(2010){Lyne}, {Hobbs}, {Kramer}, {Stairs}, \&
  {Stappers}}]{2010Sci...329..408L}
{Lyne}, A., {Hobbs}, G., {Kramer}, M., {Stairs}, I., \& {Stappers}, B. 2010,
  Science, 329, 408

\bibitem[{{Manchester}(2018)}]{2018IAUS..337..197M}
{Manchester}, R.~N. 2018, in Pulsar Astrophysics the Next Fifty Years, ed.
  P.~{Weltevrede}, B.~B.~P. {Perera}, L.~L. {Preston}, \& S.~{Sanidas}, Vol.
  337, 197--202

\bibitem[{{Melatos} \& {Warszawski}(2009)}]{2009ApJ...700.1524M}
{Melatos}, A. \& {Warszawski}, L. 2009, \apj, 700, 1524

\bibitem[{{Palfreyman} {et~al.}(2018){Palfreyman}, {Dickey}, {Hotan},
  {Ellingsen}, \& {van Straten}}]{2018Natur.556..219P}
{Palfreyman}, J., {Dickey}, J.~M., {Hotan}, A., {Ellingsen}, S., \& {van
  Straten}, W. 2018, \nat, 556, 219

\bibitem[{{Parthasarathy} {et~al.}(2019){Parthasarathy}, {Shannon}, {Johnston},
  {Lentati}, {Bailes}, {Dai}, {Kerr}, {Manchester}, {Os{\l}owski}, {Sobey},
  {van Straten}, \& {Weltevrede}}]{2019MNRAS.489.3810P}
{Parthasarathy}, A., {Shannon}, R.~M., {Johnston}, S., {et~al.} 2019, \mnras,
  489, 3810

\bibitem[{{Rankin}(1990)}]{1990ApJ...352..247R}
{Rankin}, J.~M. 1990, \apj, 352, 247

\bibitem[{{Ransom}(2011)}]{2011ascl.soft07017R}
{Ransom}, S. 2011, {PRESTO: PulsaR Exploration and Search TOolkit},
  Astrophysics Source Code Library, record ascl:1107.017

\bibitem[{{Ransom} {et~al.}(2003){Ransom}, {Cordes}, \&
  {Eikenberry}}]{2003ApJ...589..911R}
{Ransom}, S.~M., {Cordes}, J.~M., \& {Eikenberry}, S.~S. 2003, \apj, 589, 911

\bibitem[{{Rasmussen} \& {Williams}(2006)}]{2006gpml.book.....R}
{Rasmussen}, C.~E. \& {Williams}, C. K.~I. 2006, {Gaussian Processes for
  Machine Learning}

\bibitem[{{Scargle}(1982)}]{1982ApJ...263..835S}
{Scargle}, J.~D. 1982, \apj, 263, 835

\bibitem[{{Shang} {et~al.}(2020){Shang}, {Zhi}, {Dang}, \&
  {Wang}}]{2020Ap&SS.365...70S}
{Shang}, L.~H., {Zhi}, Q.~J., {Dang}, S.~J., \& {Wang}, Q.~S. 2020, \apss, 365,
  70

\bibitem[{{Shaw} {et~al.}(2022{\natexlab{a}}){Shaw}, {Mickaliger}, {Stappers},
  {Lyne}, {Keith}, {Weltevrede}, \& {Basu}}]{2022ATel15622....1S}
{Shaw}, B., {Mickaliger}, M.~B., {Stappers}, B.~W., {et~al.}
  2022{\natexlab{a}}, The Astronomer's Telegram, 15622, 1

\bibitem[{{Shaw} {et~al.}(2022{\natexlab{b}}){Shaw}, {Stappers}, {Weltevrede},
  {Brook}, {Karastergiou}, {Jordan}, {Keith}, {Kramer}, \&
  {Lyne}}]{2022MNRAS.513.5861S}
{Shaw}, B., {Stappers}, B.~W., {Weltevrede}, P., {et~al.} 2022{\natexlab{b}},
  \mnras, 513, 5861

\bibitem[{{Skilling}(2004)}]{2004AIPC..735..395S}
{Skilling}, J. 2004, in American Institute of Physics Conference Series, Vol.
  735, Bayesian Inference and Maximum Entropy Methods in Science and
  Engineering: 24th International Workshop on Bayesian Inference and Maximum
  Entropy Methods in Science and Engineering, ed. R.~{Fischer}, R.~{Preuss}, \&
  U.~V. {Toussaint} (AIP), 395--405

\bibitem[{{Weltevrede} {et~al.}(2011){Weltevrede}, {Johnston}, \&
  {Espinoza}}]{2011MNRAS.411.1917W}
{Weltevrede}, P., {Johnston}, S., \& {Espinoza}, C.~M. 2011, \mnras, 411, 1917

\bibitem[{{Yan} {et~al.}(2020){Yan}, {Manchester}, {Wang}, {Wen}, {Yuan},
  {Lee}, \& {Chen}}]{2020MNRAS.491.4634Y}
{Yan}, W.~M., {Manchester}, R.~N., {Wang}, N., {et~al.} 2020, \mnras, 491, 4634

\bibitem[{{Yu} {et~al.}(2013){Yu}, {Manchester}, {Hobbs}, {Johnston}, {Kaspi},
  {Keith}, {Lyne}, {Qiao}, {Ravi}, {Sarkissian}, {Shannon}, \&
  {Xu}}]{2013MNRAS.429..688Y}
{Yu}, M., {Manchester}, R.~N., {Hobbs}, G., {et~al.} 2013, \mnras, 429, 688

\bibitem[{{Zhang} \& {Xie}(2012)}]{2012ApJ...761..102Z}
{Zhang}, S.-N. \& {Xie}, Y. 2012, \apj, 761, 102

\bibitem[{{Zhao} {et~al.}(2017){Zhao}, {Ng}, {Lin}, {Takata}, {Cai}, {Hu},
  {Yen}, {Tam}, {Hui}, {Kong}, \& {Cheng}}]{2017ApJ...842...53Z}
{Zhao}, J., {Ng}, C.~W., {Lin}, L.~C.~C., {et~al.} 2017, \apj, 842, 53

\bibitem[{{Zhou} {et~al.}(2023){Zhou}, {G{\"u}gercino{\u{g}}lu}, {Yuan}, {Ge},
  {Yu}, {Zhang}, {Zhang}, {Feng}, \& {Ye}}]{2023MNRAS.519...74Z}
{Zhou}, S.~Q., {G{\"u}gercino{\u{g}}lu}, E., {Yuan}, J.~P., {et~al.} 2023,
  \mnras, 519, 74

\bibitem[{{Zhou} {et~al.}(2024){Zhou}, {Ye}, {Ge}, {G{\"u}gercino{\u{g}}Lu},
  {Zheng}, {Yu}, {Yuan}, \& {Zhang}}]{2024arXiv240809204Z}
{Zhou}, S.~Q., {Ye}, W.~T., {Ge}, M.~Y., {et~al.} 2024, arXiv e-prints,
  arXiv:2408.09204

\bibitem[{{Zhou} {et~al.}(2022){Zhou}, {Wang}, {Wang}, {Yuan}, {Kou}, \&
  {Dang}}]{Zhou2022}
{Zhou}, Z.-R., {Wang}, J.-B., {Wang}, N., {et~al.} 2022, Research in Astronomy
  and Astrophysics, 22, 095008

\bibitem[{{Zubieta} {et~al.}(2022){Zubieta}, {Del Palacio}, {Garcia}, {Gancio},
  {Lousto}, {Combi}, {Combi}, {Gutierrez}, {Lopez-Armengol}, {Simaz Bunzel}, \&
  {Sosa-Fiscella}}]{2022ATel15638....1Z}
{Zubieta}, E., {Del Palacio}, S., {Garcia}, F., {et~al.} 2022, The Astronomer's
  Telegram, 15638, 1

\bibitem[{{Zubieta} {et~al.}(2024){Zubieta}, {Garc{\'\i}a}, {del Palacio},
  {Araujo Furlan}, {Gancio}, {Lousto}, {Combi}, \&
  {Espinoza}}]{2024A&A...689A.191Z}
{Zubieta}, E., {Garc{\'\i}a}, F., {del Palacio}, S., {et~al.} 2024, \aap, 689,
  A191

\bibitem[{{Zubieta} {et~al.}(2023){Zubieta}, {Missel}, {Sosa Fiscella},
  {Lousto}, {del Palacio}, {L{\'o}pez Armengol}, {Garc{\'\i}a}, {Combi},
  {Wang}, {Combi}, {Gancio}, {Negrelli}, \&
  {Guti{\'e}rrez}}]{2023MNRAS.521.4504Z}
{Zubieta}, E., {Missel}, R., {Sosa Fiscella}, V., {et~al.} 2023, \mnras, 521,
  4504

\end{thebibliography}
%

\end{document}